\documentclass{naturetw}
\usepackage{graphicx}
\usepackage{amsmath,amssymb}
\usepackage{caption}
\usepackage[tight]{units}
\usepackage{textcomp}

\usepackage{hyperref}
\hypersetup{
    bookmarks=true,         
    unicode=false,          
    pdftoolbar=true,        
    pdfmenubar=true,        
    pdffitwindow=false,     
    pdftitle={},    
    pdfauthor={Tobias Witting},     
    pdfsubject={},   
    pdfcreator={Tobias Witting},   
    pdfproducer={me},  
    pdfkeywords={attosecond, OPCPA, streaking}, 
    pdfnewwindow=true,      
    colorlinks=false,       
    linkcolor=red,          
    citecolor=green,        
    filecolor=magenta,      
    urlcolor=cyan           
  }

\title{Generation and characterisation of isolated attosecond pulses
   at 100\,kHz repetition rate}

\author{Tobias Witting$^{1,*}$, Mikhail Osolodkov$^1$, Felix
  Schell$^1$, Felipe Morales$^1$, Serguei Patchkovskii$^1$, Peter
  \v{S}u\v{s}njar$^1$, Fabio Cavalcante$^2$, Carmen S. Menoni$^2$,
  Claus P. Schulz$^1$, Federico J. Furch$^{1,\dagger }$ and Marc
  J. J. Vrakking$^1$}

\begin{document}

\maketitle

\begin{affiliations}
\item Max Born Institute for Nonlinear Optics and Short Pulse
  Spectroscopy, Max-Born-Strasse 2a, 12489 Berlin, Germany
\item Department of Electrical and Computer Engineering, Colorado
  State University, Fort Collins, CO 80523, USA
\item[*] email: tobias.witting@mbi-berlin.de
\item[$\dagger$] email: furch@mbi-berlin.de
\end{affiliations}

\begin{abstract}
  The generation of coherent light pulses in the extreme ultraviolet
  (XUV) spectral region with attosecond pulse durations
  constitutes the foundation of the field of attosecond
  science~\cite{krausz_attosecond_2009}. Twenty years after the first
  demonstration of isolated attosecond pulses~\cite{Hentschel}, they
  continue to be a unique tool enabling the observation and control of
  electron dynamics in atoms, molecules and
  solids~\cite{Krausz_NatPhot_8_2014, Lepine_attosecond_2014}. It has
  long been identified that an increase in the repetition rate of
  attosecond light sources is necessary for many applications in
  atomic and molecular
  physics~\cite{Ullrich,boguslavskiy_multielectron_2012}, surface
  science~\cite{cavalieri_attosecond_2007}, and
  imaging~\cite{chew_peem_2012}.
  Although high harmonic generation (HHG) at repetition rates
  exceeding \unit[100]{kHz}, showing a continuum in the cut-off region
  of the XUV spectrum was already demonstrated in 2013~\cite{Krebs},
  the number of photons per pulse was insufficient to perform pulse
  characterisation via attosecond
  streaking~\cite{Itatani_PRL_88_2002}, let alone to perform a
  pump-probe experiment.
  Here we report on the generation and full characterisation of XUV
  attosecond pulses via HHG driven by near-single-cycle pulses at a
  repetition rate of 100\,kHz. The high number of $10^6$ XUV photons
  per pulse on target enables attosecond electron streaking experiments
  through which the XUV pulses are determined to consist of a dominant
  single attosecond pulse. These results open the door for attosecond
  pump-probe spectroscopy studies at a repetition rate one or two
  orders of magnitude above current implementations.
\end{abstract}


Extreme ultraviolet (XUV) light pulses generated via high-order
harmonic generation (HHG)~\cite{Ferray_1988}, consisting of attosecond
pulse trains~\cite{Corkum,Antoine} or isolated attosecond
pulses~\cite{Hentschel}, have led to the establishment of the field of
attosecond science~\cite{krausz_attosecond_2009}. The HHG process can
be described by a very intuitive three-step model~\cite{Corkum_3step},
in which the strong electric field of the laser interacting with a gas
of atoms first induces tunnel ionisation. The created electron wave
packet is accelerated by the electric field of the laser, first away
and then back towards its parent ion, and then recombines with the
parent ion, releasing the gained kinetic energy in the form of high
energy photons at odd multiples of the driving laser frequency. Since
the process is repeated every half-cycle of the laser for which the
peak intensity is sufficient to ionise the atom, HHG naturally gives
rise to a train of attosecond pulses. As the duration of the driving
pulse is reduced, the number of pulses in the attosecond pulse train
(APT) also decreases. If the driving pulse duration approaches a
single oscillation of the optical field, and if it is possible to
control the carrier-envelope phase (CEP) of the driving pulses, the
XUV emission can be confined to a single
half-cycle~\cite{Hentschel,Sansone443,Kim,pfeifer_generating_2007}. Isolated
attosecond pulses (IAPs) generated in this manner have been
successfully implemented to observe light-induced electron tunneling
in atoms in real time~\cite{Uiberacker}, to follow the valence
electron motion in an excited ion~\cite{Goulielmakis}, and to control
electron localisation in a dissociating molecule with attosecond
precision~\cite{Sansone_H2}, among many other outstanding examples.

In all the aforementioned examples, and in the majority of
experimental setups used for attosecond science, the driving pulses
for HHG are produced by a Ti:Sapphire chirped pulse amplification
(CPA) system followed by non-linear post-compression in a gas-filled
hollow-core fibre~\cite{Nisoli:97}, which allows reaching pulse
durations near a single optical cycle ($\unit[2.67]{fs}$ at
\unit[800]{nm}) with pulse energies of at least a few
\unit[hundred]{\textmu{}J}. Typical observables in attosecond
pump-probe experiments are either the XUV absorption spectrum, or the
kinetic energy and/or angular distributions of ions or
electrons. While these schemes provide valuable information, many
research questions call for more sophisticated experimental
approaches. In particular, in experiments with atomic and molecular
targets in the gas phase it is highly attractive to implement
electron-ion coincidence detection schemes~\cite{Ullrich}, which give
access to the fully correlated three-dimensional momentum
distributions of all charged particles produced during a
photoionisation experiment. The technique requires that only one
photoionisation event is produced per laser pulse in order to avoid
the detection of false coincidences; i.e. ions and electrons that are
measured in coincidence, but that originate from different parent
atoms/molecules. In practice, this requires working at a maximum event
rate of approximately 0.2 times the repetition rate of the light
source. It follows that high repetition rate light sources are
required for these types of experiments to keep data acquisition times
manageable. So far, the implementation of coincidence detection in
attosecond experiments has been limited to setups working with
Ti:Sapphire CPAs running at
$\leq\unit[10]{kHz}$~\cite{Sabbar_RevSciInst_85_2014}, where the
technique has been successfully applied to the study of small
molecular
systems~\cite{cattaneo_attosecond_2018,vos_orientation-dependent_2018}.
However, more complex problems involving the study of low probability
ionisation or dissociation channels require higher repetition rates.
Furthermore condensed phase experiments such as photoelectron emission
microscopy (PEEM)~\cite{Stockman:07} and attosecond photoelectron
streaking~\cite{cavalieri_attosecond_2007,okell_temporal_2015} at
surfaces would also benefit from high repetition rate attosecond XUV
sources.
Since a high number of photoelectrons removed from a surface with a
single laser shot has detrimental effects on the energy
resolution through space-charge effects, it is advantageous to keep
the number of photoelectrons generated per shot low. To achieve high
signal-to-noise ratio measurements that go beyond proof of principle
experiments, increasing the laser repetition rate is the only viable
route.

Ti:Sapphire CPAs, delivering CEP-stable, ultrashort pulses with the
necessary pulse energies for attosecond pulse generation (typically up
to a few mJ) are not scalable beyond \unit[10]{kHz}. Optical
parametric chirped pulse amplifiers (OPCPA)~\cite
{Furch:16,Rothhardt:12,Prinz:15,Matyschok:13,Furch,Hrisafov:20} and
direct post-compression of high repetition rate ytterbium based
CPAs~\cite{lavenu_high-power_2019,Haedrich:20,muller_multipass_2021}
are viable alternatives capable of generating CEP-stable, few-cycle
pulses with multi-\textmu{}J-level energy, at hundreds of kHz
repetition rate. Moreover, OPCPAs have extended the range of
CEP-stable, ultrashort pulses with moderate energy and high repetition
rate towards the mid-IR
region~\cite{Pupeikis:20,Thire:18,Neuhaus:18,Elu:17,Mero:18}. Some
high repetition rate OPCPAs operating at \unit[800]{nm} have been
utilised to generate high-order harmonics with a spectrum in the XUV
range showing sensitivity to the value of the CEP of the driving
pulse~\cite{Krebs,Harth}. Krebs et al.~\cite{Krebs}
demonstrated a continuous XUV spectrum in the cut-off region using
\unit[6.6]{fs} \unit[918]{nm} pulses from an OPCPA, suggesting the
existence of an isolated pulse in this region of the XUV
spectrum. However, the number of XUV photons per pulse was not
sufficient to carry out pulse characterisation with the attosecond
streaking technique to determine the temporal structure and pulse
duration.

Here we present, to the best of our knowledge, the first demonstration
of an attosecond electron streaking experiment driven by XUV pulses
operating at \unit[100]{kHz} repetition rate, i.e. at least an order
of magnitude higher than any previous implementation. The ability to
do so stems from the available high number of $10^6$ XUV photons per
pulse on target, which is comparable to typical attosecond light
sources driven by post-compressed Ti:Sapphire CPA systems.
The temporal structure of the pulse is dominated by a single XUV pulse
with sub-\unit[140]{as} duration, demonstrating the suitability of our
source for attosecond pump-probe experiments.

\begin{figure}
  \centering
  \includegraphics[width=\linewidth]{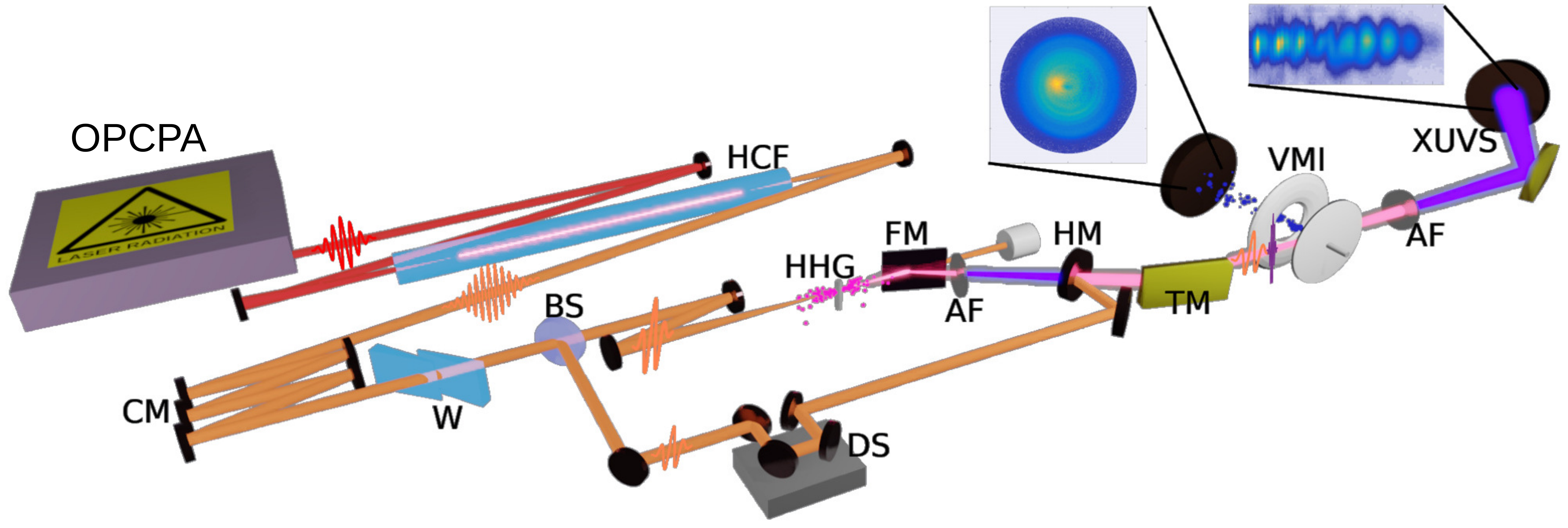}
  \caption{\textbf{$\mid$ Schematic view of the experimental
      setup}. Pulses from an OPCPA laser system are compressed in a
    hollow-core fibre (HCF). The pulses are split by a beam splitter
    (BS). HHG takes place in one arm of the interferometer.  The
    pulses are recombined with a variable delay and focused into the
    velocity map imaging spectrometer (VMI). CM: Chirped mirrors; W:
    fused silica wedges; DS: delay stage; FM: filter mirror; AF:
    aluminium filter; HM: holey mirror; TM: toroidal mirror; XUVS: XUV
    spectrometer.}
  \label{fig:main_setup}
\end{figure}

\begin{figure}
  \centering
  \includegraphics[width=\textwidth]{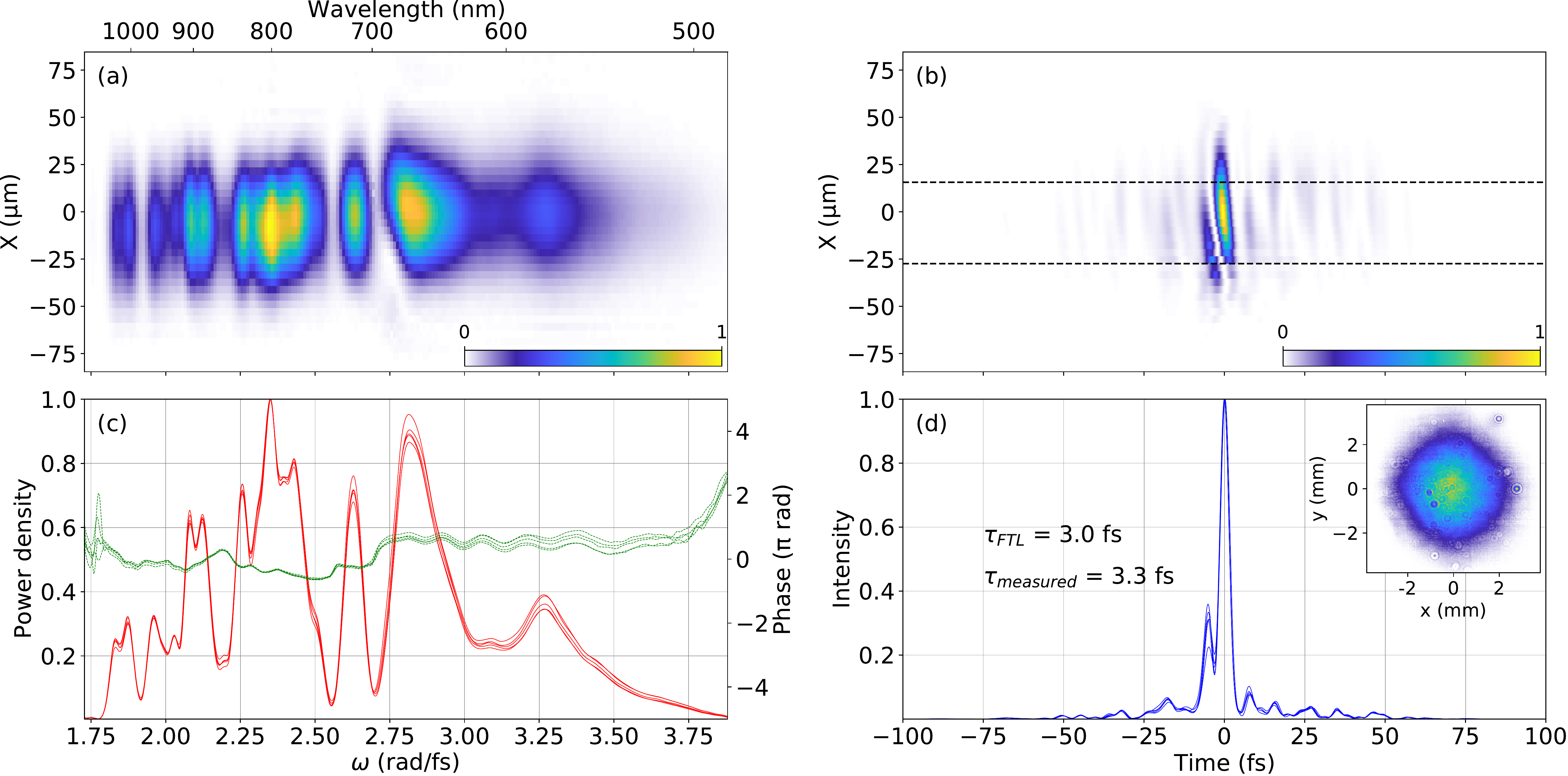}  
  \caption{\textbf{$\mid$ Spatio-temporal characterisation of the
      near-single cycle pulses}.
    (a) Spatio-spectral intensity distribution
    $|E(x, y_0, \omega)|^2$. (b) Spatio-temporal intensity
    distribution $|E(x, y_0, t)|^2$. (c) Spatially integrated spectral
    intensity (red line), and spatially integrated spectral phase
    (green line). Results from 5 consecutive measurements are
    shown. (d) Spatially integrated temporal intensity profile (blue
    line); results from 5 consecutive measurements are plotted. The
    inset in (d) shows the spatial near-field beam profile on a CMOS
    camera.}
  \label{fig:main_seafspiderhcf}
\end{figure}

A sketch of the experimental setup is shown in
Fig.~\ref{fig:main_setup}. An OPCPA delivered CEP-stable \unit[7]{fs}
pulses at a central wavelength of \unit[790]{nm} with up to
\unit[190]{\textmu{}J} of energy per pulse and at a repetition rate of
\unit[100]{kHz}~\cite{Furch,Witting_spatiotemporalOPCPA,Hoff:18}. In
HHG the \unit[7]{fs} pulses from the OPCPA system, containing
$~\unit[2.6]{}$ optical cycles, lead to the generation of short APTs
containing 7 pulses with varying intensity~\cite{Osolodkov2020}. To
enable efficient amplitude gating of IAPs, the \unit[7]{fs} OPCPA
pulses were compressed to near-single cycle pulse durations, using
hollow fibre pulse compression in a \unit[1]{m} long fibre filled with
neon gas in gradient pressure configuration. Spectra spanning more
than one octave supporting \unit[3]{fs} pulses were
achieved. Fig.~\ref{fig:main_seafspiderhcf} summarises the
spatio-temporal characterisation performed with the SEA-F-SPIDER
technique~\cite{witting_characterization_2011}, which measures the
spatially dependent field in one spatial plane, i.e. either
$E(x, y_0, t)$ or $E(x_0, y, t)$. As the non-collinear angle of the
OPCPA system lies in the horizontal $x$-plane, eventual
spatio-temporal couplings would be expected in this
plane~\cite{Witting_spatiotemporalOPCPA}. In
Fig.~\ref{fig:main_seafspiderhcf}(a) the spatio-spectral intensity
distribution $|E(\omega, x, y_0)|^2$ is shown. Apart form small
spatial variations (e.g. a frequency-dependent spot size), no
significant space-time couplings were observed.
In Fig.~\ref{fig:main_seafspiderhcf}(c) the spatially integrated
spectral intensity (red line) and spectral phase (green line) are
shown for a series of five measurements acquired over the space of one
minute, demonstrating the excellent short term stability of the
source. The central wavelength is \unit[760]{nm}. The spectral phase
is flat within the compression range of the chirped mirrors form
\unit[500]{} to \unit[1050]{nm} and shows small oscillations caused by
the chirped mirror compressor.  Fig.~\ref{fig:main_seafspiderhcf}(b)
shows the spatio-temporal intensity profile and (d) the spatially
integrated temporal intensity. The pulse duration is
$\unit[3.3\pm 0.1]{fs}$, corresponding to only 1.3 optical cycles. The
inset of Fig.~\ref{fig:main_seafspiderhcf}(d) shows the near-field
spatial profile of the beam after the HCF and chirped mirror
compressor measured by a CMOS camera.

The near-single-cycle near-infrared (NIR) pulses were sent to an
attosecond beamline sketched in Fig.~\ref{fig:main_setup} and
described in detail in~\cite{Osolodkov2020}. The pulses were split with
the majority of the pulse energy used for HHG. For this, the
near-single-cycle pulses were focused to an approximate intensity of
$2\times 10^{14}$ W/cm$^2$ inside a cylindrical gas cell with a
diameter of \unit[2]{mm}, filled with krypton at a backing pressure of
\unit[70]{mbar}. The optimum CEP was chosen by fine tuning the
compression (see \textbf{supplementary information}). After HHG the
XUV and remaining NIR radiation were separated utilising a dielectric
filter mirror and an Al filter. The smaller portion of the NIR energy
was used as probe beam. A piezo translation stage in the probe arm was
used to provide a variable delay. XUV and NIR pulses were recombined
with a holey mirror. Both beams were focused into the experimental
chamber by a gold-coated toroidal mirror. The spectrally integrated
photon flux on target for the experiment was measured with an XUV
photodiode and amounted to $10^6$ photons per shot (comparable to
typical Ti:Sapphire laser driven attosecond setups), which amounted to an
unprecedented flux of $10^{11}$ photons per second on target. In the
experimental chamber the attosecond XUV pulse ionised neon atoms in
the presence of a strong NIR probe pulse
($I\approx \unit[4.6]{TW/cm^2}$) and the resulting photoelectron
momentum distributions were measured with a velocity map imaging
spectrometer (VMI)~\cite{Ghafur}.

For the characterisation of the attosecond XUV pulses the attosecond
streaking technique~\cite{Itatani_PRL_88_2002} was employed.
Single photon ionisation by the XUV pulses launches a photoelectron
wave packet into the continuum.  In the absence of the NIR pulse, the
photoelectron wave packet reproduces the amplitude and phase of the
XUV pulses, re-shaped by the photoionisation cross section, and
shifted in energy by the ionisation potential of the target atom. The
low photon energy limit of the XUV spectrum, given by the transmission
window of the aluminium filter at approximately \unit[15.5]{eV} lies a
few eV below the ionisation threshold of neon
(\unit[21.56]{eV}). Therefore, information about the low energy end of
the XUV pulse spectrum is lost during photoionisation and the XUV
pulse that can be retrieved corresponds to the portion of the spectrum
above the ionisation potential of neon. The presence of the NIR pulse
modulates the photoelectron kinetic energy distribution depending on
the value of the NIR field vector potential at the moment of
ionisation giving rise to a streaking effect. Measurements of the
photoelectron kinetic energy distribution at a particular observation
angle as a function of XUV-NIR delay (i.e., the electron streaking
trace) contain amplitude and phase information on both, the NIR and
the XUV fields. Within the strong field approximation (SFA), the
probability amplitude for producing photoelectrons in the continuum
with final momentum $\vec{p}$ is given
by~\cite{Itatani_PRL_88_2002,Lewenstein,Mairesse}:
\begin{equation}
  a(\vec{p},\tau) = -i \int_{-\infty}^{\infty} \,dt\, d_{\vec{p}+\vec{A}(t)} E_{\text{XUV}}(t-\tau)e^{i[(W + \text{IP})t -\int_{t}^{\infty} dt'( \vec{p}\cdot\vec{A}(t')+A(t')^2/2)] }
\label{eq:trace}
\end{equation}
where $\tau$ is the time-delay between the NIR and XUV fields,
$E_{\text{XUV}}$ is the complex amplitude of the electric field of the
XUV pulse, $\vec{A}(t)$ is the vector potential of the NIR field,
$d_{\vec{p}+\vec{A}(t)}$ is the complex dipole transition matrix
element from the ground state to the continuum, $\text{IP}$ is the
ionisation potential of the ionised atom, $\vec{p}+\vec{A}(t)$ is the
instantaneous kinetic momentum, and $W = p^{2}/2$ is the measured
photoelectron kinetic energy. Atomic units are used. The measured
electron yield in a streaking trace is proportional to
$|a(\vec{p},\tau)|^2$.

\begin{figure}
  \centering
  \includegraphics[width=\textwidth]{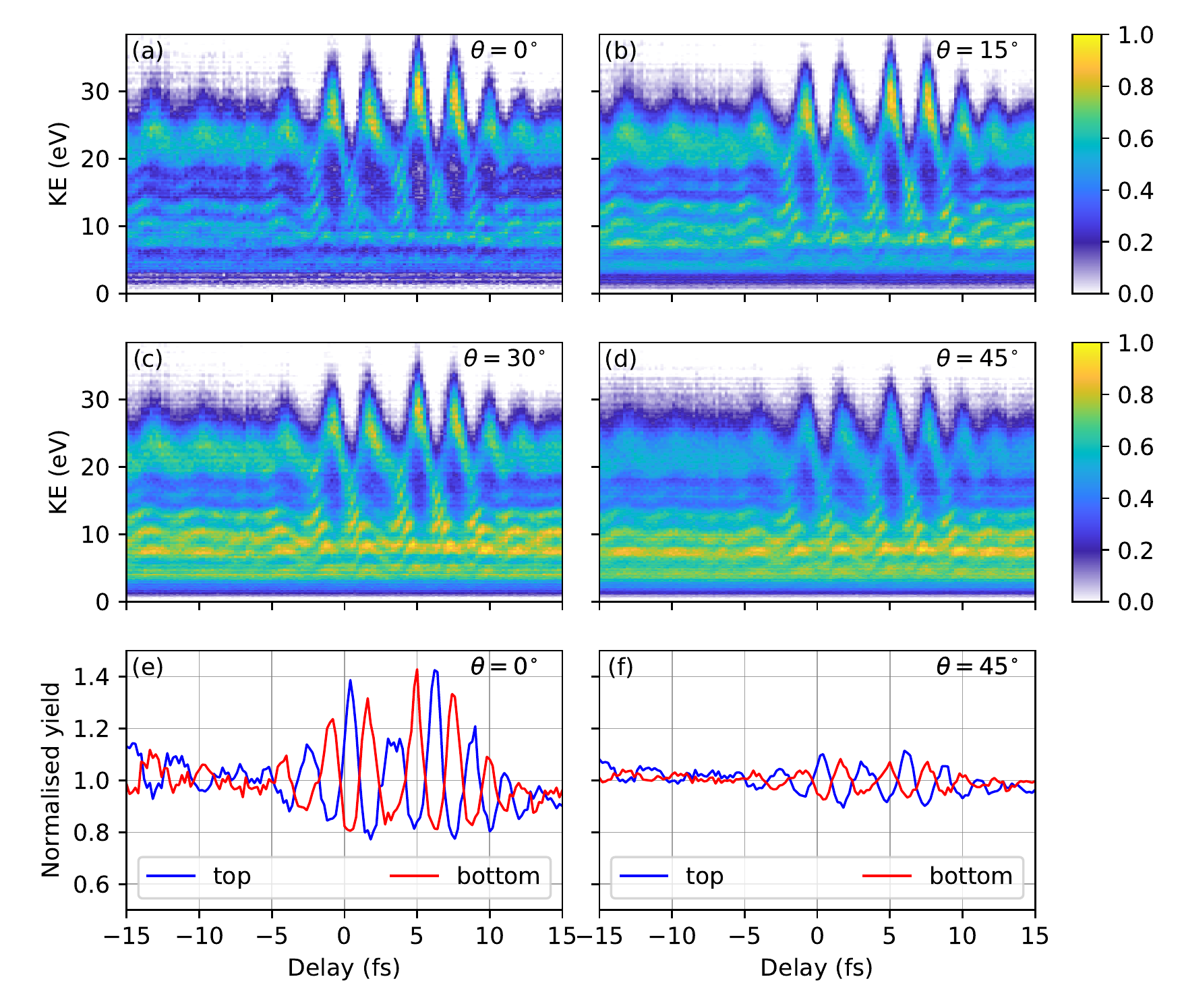}
  \caption{\textbf{$\mid$ Attosecond electron streaking traces}.
    Streaking traces from photoelectron kinetic energy distributions
    at different observation angles $\theta$ with respect to the
    laser polarisation. (a) $\theta = 0\,^{\circ}$, (b)
    $\theta = 15\,^{\circ}$, (c) $\theta = 30\,^{\circ}$, and (d)
    $\theta = 45\,^{\circ}$. In all cases the colour scale indicates
    the electron yield and it is normalised to the maximum of the
    trace. (e) and (f) show normalised electron yields as a function
    of delay for observation angles of $0\,^{\circ}$ and
    $45\,^{\circ}$ respectively, and computed in the top and bottom
    parts of the detector.}
  \label{fig:main_traces}
\end{figure}

Streaking traces for a range of angles with respect to the
polarisation vector were constructed from the measured photoelectron
momentum distributions (see \textbf{methods}
section). Figs.~\ref{fig:main_traces} (a)-(d) show the photoelectron
kinetic energy distributions as a function of XUV-NIR delay, taken at
observation angles of 0\,$^{\circ}$, 15\,$^{\circ}$, 30\,$^{\circ}$,
and 45\,$^{\circ}$, respectively. The electron streaking traces
clearly show the modulation of the photoelectron kinetic energy
distributions as a result of the presence of the NIR field during
ionisation. The spectral modulations with an energy spacing of
approximately twice the NIR photon energy $\omega_{\mathrm{NIR}}$ at
photon energies below $\approx \unit[20]{eV}$ indicates the presence
of satellite pulses at a delay of $\pm T_{\mathrm{NIR}}/2$ from the main
XUV pulse. At the same time, the smooth spectrum above
$\approx \unit[20]{eV}$ kinetic energy suggests that a clean isolated
attosecond pulse exists in this portion of the XUV spectrum.

In order to retrieve the XUV and NIR pulses from the experimental
trace, most retrieval methods replace the product
$\vec{p}\cdot\vec{A}(t)$ in equation~\eqref{eq:trace} by
$\vec{p_o}\cdot\vec{A}(t)$, with $\vec{p_o}$, the central momentum of
the kinetic energy distribution. This central momentum approximation
(CMA), leads to an overestimation of the strength of the streaking
effect at low kinetic energies. Validity of the CMA requires the
central momentum of the kinetic energy distribution to be considerably
larger than its bandwidth ($|\vec{p_o}| >> \Delta p$), a condition
that is certainly not met in the measurements shown in
Fig.~\ref{fig:main_traces}. Therefore, the Volkov-transform
generalized projection algorithm (VTGPA)~\cite{Keathley} was chosen as
retrieval method to analyse the experimental data. The VTGPA method
determines an explicit solution for $E_{\text{XUV}}(t)$ by finding a
local minimum of an error function (or a figure of merit
expression)~\cite{Keathley}. If the error function for a particular
streaking trace contains many local minima, the solution of the VTGPA
may converge to any of those minima. Therefore, to complement the
analysis of our data sets by the VTGPA, an alternative method was
developed, based on a global optimisation routine. This method, which
we call streaking global optimisation (SGO) guesses the XUV spectrum
from a weakly streaked region of the experimental trace, and utilises
equation~\eqref{eq:trace} and standard global optimisation routines to
fit the spectral phase. The attosecond pulse retrieval algorithms and
the data processing are discussed in detail in the
\textbf{supplementary information}.

Note that equation~\eqref{eq:trace} assumes that the effect of the
attractive Coulomb field on the escaping photoelectron can be
neglected, and furthermore, that the only ionisation pathway is single
photon ionisation by the XUV field. As a result of the second
assumption and due to the phase term $\vec{p}\cdot\vec{A}(t)$, it is
expected that electron streaking traces at different observation
angles differ in the strength of the streaking effect, as observed in
the traces shown in Figs.~\ref{fig:main_traces} (a) to (d).
Additional differences in the shape of the photoelectron spectra arise
due to the angle-dependence of the dipole transition matrix
element. In Figs.~\ref{fig:main_traces} (a) to (d) the relative
spectral amplitude at lower photoelectron energies increases with
angle.  In addition, inspection of the energy-integrated electron
yield as a function of the XUV-NIR delay shows a strong modulation of
the yield in the regions where the streaking field is strongest. This
modulation has an opposite sign in the top and bottom portions of the
streaking trace.
Due to the strong NIR field and very short pulse duration one can
expect a high degree of ground state polarisation by the strong NIR
field, which in turn strongly changes the photoionisation probability.

%

In order to asses the validity of the VTGPA and the SGO in the
presence of the aforementioned effects in the experimental data, the
algorithms were first applied to simulated streaking traces, that
where obtained by numerically solving the time-sependent
Schr\"{o}dinger equation (TDSE) in the single-active electron
approximation, with an effective potential for
neon~\cite{Serguei_Ne}. Analysis of the simulated results shows that
both methods are able to satisfactorily retrieve the input XUV field
with a relative pulse field error for normalised
fields~\cite{Dorrer_rmsf} on the order of \unit[0.1]{}. Further
details can be found in the \textbf{supplementary information}.

The VTGPA and SGO methods were applied to a set of experimental traces
at different observation angles. In total, 32 different experimental
traces were analysed with each retrieval method. More details about
the data processing and the analysis are presented in the
\textbf{supplementary information}. Fig.~\ref{fig:main_average_pulses}
shows a summary of the results. Figs.~\ref{fig:main_average_pulses}
(a) and (c) show the absolute value squared of the average XUV pulse
envelope retrieved with the VTGPA and the SGO algorithms,
respectively. A confidence region of $\pm 1 \sigma$ is indicated by
the light blue shaded area, which is not visible on a linear scale.
The pulse retrievals by the VTGPA result in an average pulse duration
of $\unit[132\pm 5]{as}$ FWHM, while the SGO algorithm retrieves XUV
pulses with a FWHM of $\unit[124\pm 3]{as}$. The insets in (a) and (c)
show the temporal intensity profiles on a logarithmic scale. The
pre-pulse satellite located at $\approx\unit[-1.35]{fs}$
(approximately one half of the NIR laser period) has a relative
intensity of $0.7 \times 10^{-3}$, and $1.0 \times 10^{-3}$ for the
VTGPA and SGO retrievals respectively. The post-pulse satellite
located at $\approx\unit[+1.35]{fs}$ has a relative intensity of
$4 \times 10^{-4}$, and $6 \times 10^{-4}$ for the VTGPA and SGO
retrievals respectively. The satellite pre- and
post-pulses stem from XUV emission from the two weaker neighbouring
half-cycles of the NIR field and are not unexpected for purely
amplitude gated HHG. Their extremely low relative intensity
underscores the usefulness of our source for attosecond pump-probe
experiments. Figs.~\ref{fig:main_average_pulses} (b) and (d) show the
retrieval results in the spectral domain. The mean of the retrieved
spectral intensities is plotted as a dark green line with the
confidence region of $\pm 1\sigma$ indicated by the light green shaded
area. Similarly, the means of the spectral phases are plotted as dark
red dashed lines, with the $\pm 1\sigma$ confidence intervals as light
red shaded areas. In both cases the phases are essentially flat with a
residual negative third order component, as is expected for XUV pulses
near the cut-off region with phase-matching favouring the short
trajectories~\cite{varju_physics_2009}. Individual reconstructed
streaking traces are shown in the \textbf{supplementary information}.

In summary, we have demonstrated the generation and characterisation
of high flux isolated attosecond XUV pulses at an unprecedented
repetition rate of \unit[100]{kHz}. This was achieved by direct
amplitude gating of HHG using near-single-cycle (\unit[3.3]{fs})
driving pulses obtained by hollow fibre compression of pulses from an
OPCPA laser system.
The high XUV
photon flux of our source allowed performing attosecond electron
streaking experiments. The measured streaking traces clearly indicate
the presence of isolated XUV pulses, which have been retrieved
utilising two different retrieval algorithms, previously tested by
analysing simulated streaking traces generated by numerically solving
the time-dependent Schr\"{o}dinger equation. It is envisioned that
this high repetition rate source of attosecond pulses will enable
attosecond pump-probe spectroscopy studies with electron-ion
coincidence detection, allowing data collection speeds one order of
magnitude higher than currently operating systems.

\begin{methods}



\subsection{High repetition rate near single-cycle laser.}
The pulses from the OPCPA
system~\cite{Furch,Witting_spatiotemporalOPCPA,Hoff:18} were sent
into a \unit[1]{m} long \unit[340]{\textmu{}m} diameter hollow-core
fibre (HCF). The HCF was operated in differential pumping
configuration~\cite{Suda,witting_characterization_2011,okell_carrier-envelope_2013},
i.e. the entrance side was kept at a
$\approx\unit[1\times{}10^{-1}]{mbar}$ vacuum, whilst the exit volume
was filled with neon gas at a pressure of 2.5 bar.
Dispersion compensation was achieved using chirped mirrors (Ultrafast
Innovations, PC70) with a negative group delay dispersion in the range
of 500 to \unit[1050]{nm}. A pair of ultra-thin fused silica wedges
was employed for fine tuning of the dispersion to achieve optimal
compression inside the HHG target.
The pulse energy after losses in the chirped mirror compressor and the
ultra-thin wedges was \unit[95]{\textmu{}J}.

\subsection{Streaking traces.}
For each delay between the XUV and NIR pulses, a VMI image was
acquired with an integration time of approximately \unit[15]{seconds}.
Three-dimensional momentum distributions were retrieved from the
experimental images by applying an Abel
inversion~\cite{hickstein_direct_2019} based on the rBASEX
algorithm~\cite{ryazanov_development_2012,ryazanov_improved_2013}. This results in angular
distributions described by an expansion in terms of Legendre
polynomials. Legendre terms up to order 14 were used to account for
the maximum number of photons involved in the strong NIR field dressed
photoionisation process.
The Legendre expansion allows building streaking traces at any
particular angle with respect to the polarisation axis. For further
analysis, streaking traces at angles of 0, 15, 30, 45, 135, 150, 165,
and 180 degrees with respect to the laser polarisation axis were
constructed. A separate experimental image of photoionisation of neon
by harmonics produced with narrowband XUV pulses obtained by
evacuating the hollow fibre was used to calibrate the kinetic energy
axis of the VMI spectrometer.

\end{methods}



\newpage


\begin{addendum}
\item This work was supported financially by the European Union
  Horizon 2020 Marie Curie ITN project ASPIRE (674960). C.S.M.
  acknowledges support of grant DoD ONR N00014-17-1-2536. The authors
  thank R. Peslin, A. Loudovici, and Ch. Reiter for technical support.
\item[Author contributions] T.W., M.O., F.S., P.S., C.P.S, and
  F.J.F. built the experimental setup.  F.C. and C.S.M. designed and
  manufactured the XUV filtering mirror. T.W., M.O., and
  F.J.F. performed the experiments. T.W. and F.J.F. analysed the data.
  F.M. and S.P. performed the numerical simulations.  T.W. and F.J.F wrote the
  manuscript with contributions from M.O., F.M., C.P.S., and
  M.J.J.V. All authors discussed the results and the analysis of the
  data.
\item[Additional Information] Correspondence and requests for
  materials should be addressed to T.W. or F.J.F

\item[Competing Interests] The authors declare no competing financial
  interests.
\end{addendum}


\clearpage

\clearpage

\clearpage

\clearpage

\begin{figure}
  \centering
  \includegraphics[width=\textwidth]{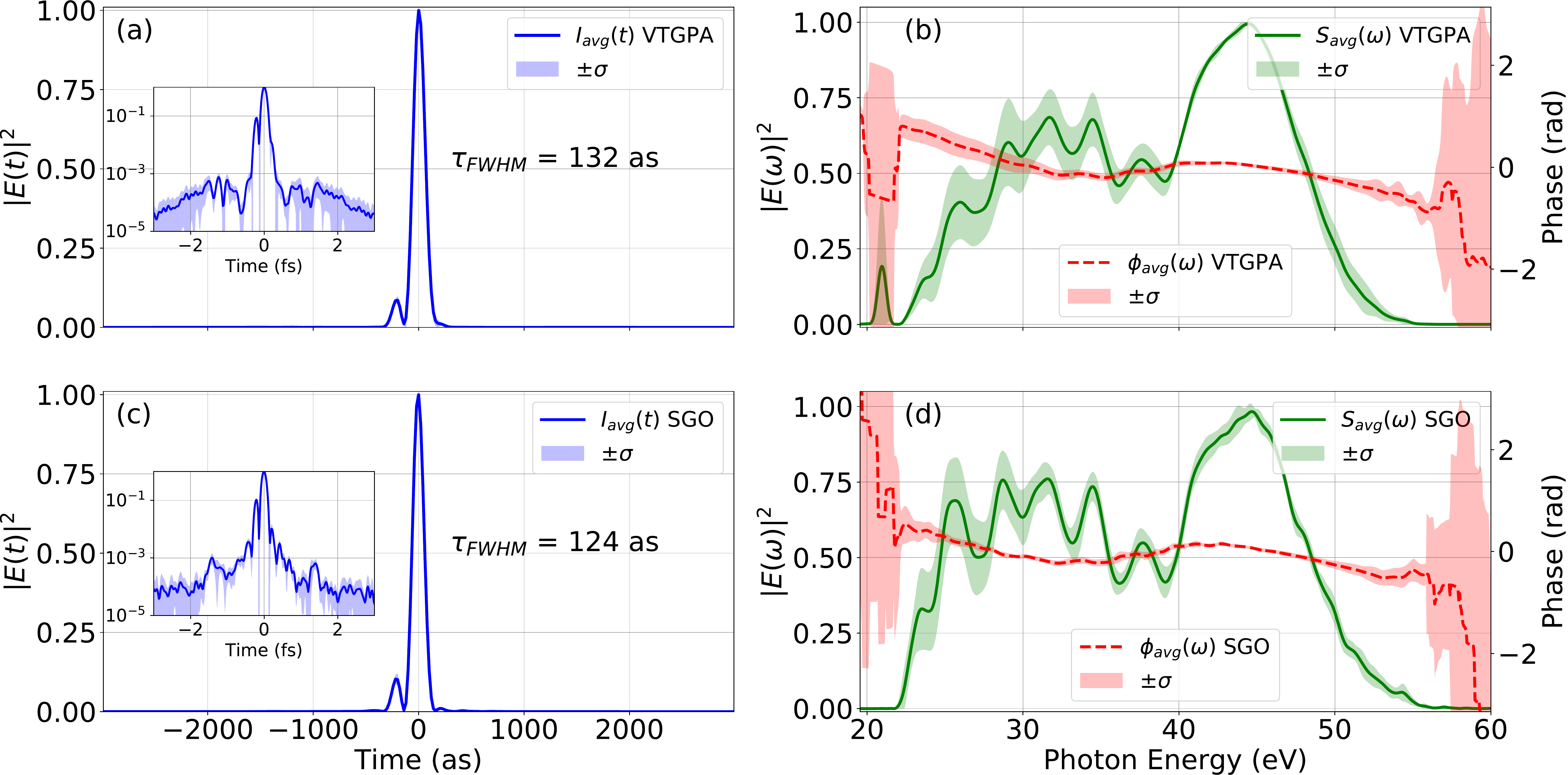}
  \caption{\textbf{$\mid$ Retrieval of the XUV IAPs}. (a) Retrieved
    XUV pulse intensity envelope normalised to its maximum value. The
    blue line corresponds to the average over all VTGPA retrievals and
    the shaded area corresponds to 1 standard deviation around the
    mean (not visible on the linear scale). The inset shows the
    intensity envelope on a logarithmic scale. (b) Retrieved average
    spectrum (green) and spectral phase (red dashed line)
    corresponding to the pulse in (a). The shaded areas correspond to
    1 standard deviation around the means. (c) and (d) show the
    corresponding results found with the SGO retrievals.}
  \label{fig:main_average_pulses}
\end{figure}

\end{document}



\maketitle

\begin{affiliations}
\item Max Born Institute for Nonlinear Optics and Short Pulse
  Spectroscopy, Max-Born-Strasse 2a, 12489 Berlin, Germany
\item Department of Electrical and Computer Engineering, Colorado
  State University, Fort Collins, CO 80523, USA
\item[*] email: tobias.witting@mbi-berlin.de
\item[$\dagger$] email: furch@mbi-berlin.de
\end{affiliations}

\section{Optimisation of the NIR pulse duration and the CEP}

\begin{figure}
  \centering
  \includegraphics[width=\textwidth]{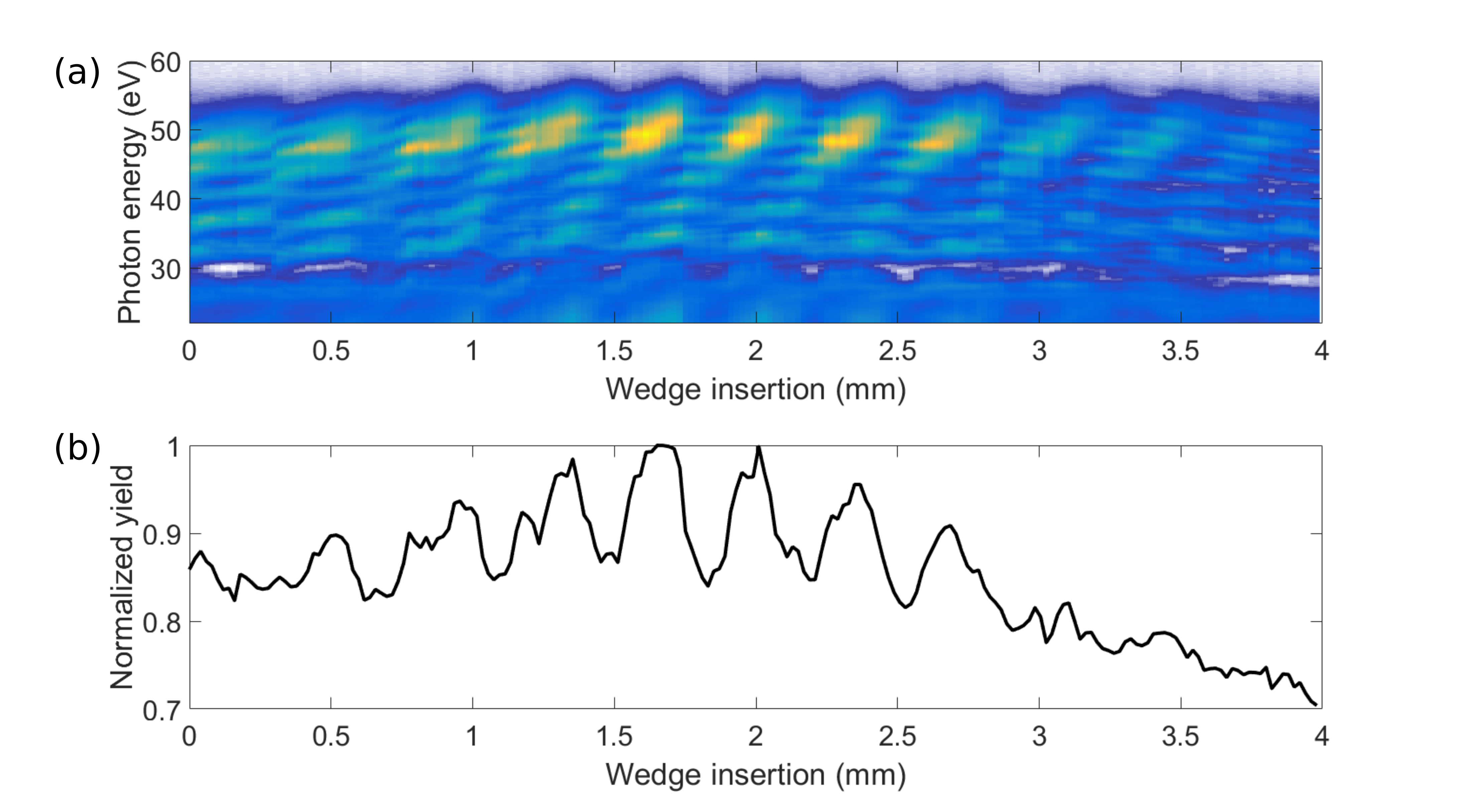}
  \caption{\textbf{$\mid$ Wedge insertion scan for NIR pulse duration
      optimisation and CEP fine-tuning}. (a) XUV spectra as a function
    of thin-wedge insertion into the NIR beam path.  (b) Normalized
    XUV photon yield as function of the wedge insertion.}
  \label{fig:suppl_wedge_scan}
\end{figure}

In order to optimise the NIR pulse duration and to find the optimum
CEP for efficient gating of one attosecond pulse a wedge scan was
performed. The amount of fused silica inserted into the path of the
NIR beam was controlled with a pair of thin fused-silica wedges
mounted on linear stages. Fig.~\ref{fig:suppl_wedge_scan} shows
measured HHG spectra as a function of the glass insertion (top) and an
integration over the spectral coordinate (bottom). From the spectrally
integrated data the optimum pulse compression for which the XUV
emission reaches a maximum cut-off can be determined. Around this
maximum in XUV cut-off, the wedge position was fine-tuned to find the
optimum CEP which minimises the spectral modulation at high photon
energy, which in turn is an indication for isolated attosecond
pulses~\cite{Baltuska}.

\section{Attosecond pulse retrieval algorithms}
\label{attopulseretrievalalgorithm}
In our implementation of the VTGPA, a first guess of the temporal
shape of the vector potential $\vec{A}(t)$ was extracted directly from
the experimental streaking trace by integrating the data over the
kinetic energy axis for energies above \unit[14]{eV}, while the
initial amplitude was set to \unit[5]{TW/cm$^2$}. The amplitude of the
vector potential was a fit-parameter during the successive
iterations. The algorithm completed 250 iterations or was terminated
when the figure of merit $\epsilon_{\text{trace}}$ (defined in
\eqref{eq:err} below) of the fit had a relative change from one
iteration to the next below $5\times10^{-5}$. The algorithm was
initialised with an XUV spectrum guessed from the experimental trace,
based on a weakly streaked region of the trace, and corrected with the
energy-dependent cross section of single photon ionization of neon.
  
In the SGO method, the initial guess of the NIR pulse and the XUV
spectrum was generated in the same way as in our implementation of the
VTGPA. The XUV spectral phase was parametrised by 70 points
equidistant on the frequency axis, in the region where the XUV
spectrum was non-zero ($\approx
\unit[21]{eV}-\unit[55]{eV}$). Subsequently, a cubic spline
interpolation was performed to build the spectral phase on the same
frequency axis as the XUV spectrum. Upon a Fourier transform, the XUV
field in the time domain and the guessed NIR pulse were used to
calculate a guessed streaking trace according to eq. (1) of the main
manuscript. This calculated trace was compared to the experimental
trace in order to obtain an error or figure of merit according to:
\begin{equation}
  \epsilon_{\text{trace}} = \frac{\sum_{i,j} |ST[i,j] - RT[i,j]|}{ \sum_
    {i,j}ST[i,j]}
  \label{eq:err}
\end{equation}
\noindent where $ST[i,j]$ and $RT[i,j]$ are two-dimensional arrays
representing the measured streaking trace and the reconstructed
streaking trace utilising the guessed pulses, respectively. The 71
resulting parameters (70 for the XUV spectral phase, 1 for the NIR
amplitude) were fit while minimising the error function utilising
standard global optimisation routines from the scipy python
package. In particular the basin-hopping
algorithm~\cite{wales_global_1997} (50 iterations) was used in order
to reach a global solution.



Beyond the points discussed in the previous section, a remaining
challenge in the retrieval process is systemic to all spectrographic
pulse retrieval
methods~\cite{walmsley_characterization_2009,stibenz_advanced_2006,escoto_advanced_2018}. The
phase of the unknown pulse is encoded in amplitudes of the measured
trace; e.g. for an unknown pulse with a satellite, the traces
corresponding to slight changes in the relative phase between main-
and satellite pulses are very similar, thus leading to comparable
values of $\epsilon_{\text{trace}}$. As a result, the algorithms tend
to stagnate in local minima of the error function
$\epsilon_{\text{trace}}$.


\section{Attosecond pulse retrievals from simulated traces}
\label{sec:pulseretrievalsfromsimulateddata}

In this section, numerical simulations are utilised to evaluate the
performance of the retrieval algorithms.  Neither the Volkov-transform
generalised projection algorithm (VTGPA)~\cite{Keathley} nor the
streaking global optimisation (SGO) take into account the effect of
the Coulomb field on the slower photoelectrons, contributions to the
electron yield from ionised excited states, higher order processes in
the two-colour photoionisation process. All these processes can alter
the observed photoelecton spectra and therefore the streaking trace,
but are not taken into account by the SFA description underlying both
the VTGPA, and SGO algorithms.
%
\begin{figure}[htbp]
  \centering
  \includegraphics[width=\textwidth]{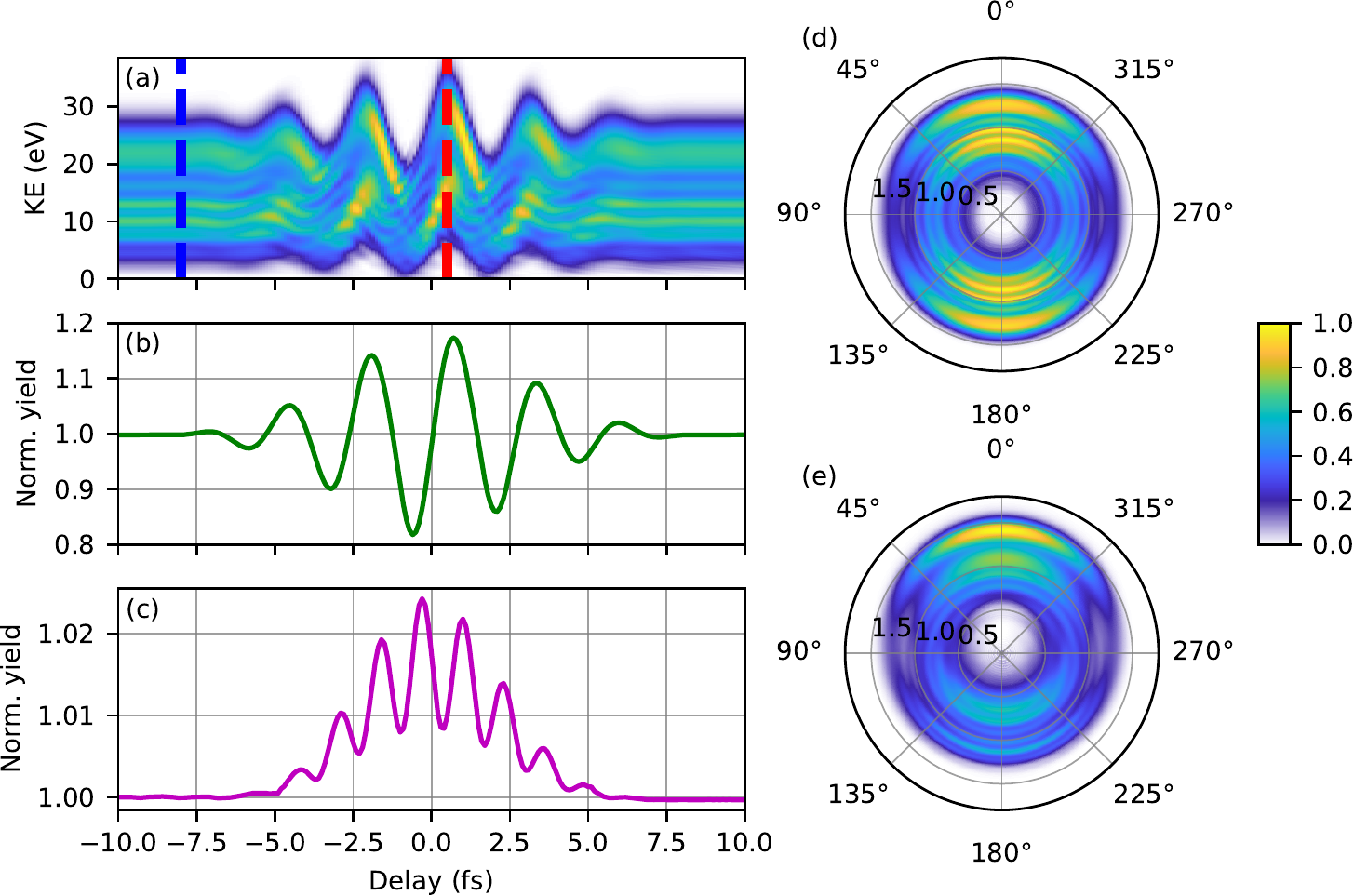}
  \caption{\textbf{$\mid$ Photoelectron yields in the streaking
      traces}. (a) Streaking trace from TDSE calculations; (b)
    Photoelectron yield calculated by integrating the trace in (a)
    over the kinetic energy axis; (c) Total yield calculated by
    integration over the full angular distribution. (d) Photoelectron
    momentum distribution at zero NIR vector potential (delay position
    indicated by the blue dashed line in (a)). (e) Photoelectron
    momentum distribution at large NIR vector potential (delay
    position indicated by the red bashed line in (a)). The colourbar
    indicating the normalised photoelectron yield is common to (a),
    (d), and (e).}
\label{fig:suppl_tdse_yield}
\end{figure}

Fig.~\ref{fig:suppl_tdse_yield}\,(a) shows a calculated streaking
trace taking into account only photoelectrons emitted at
\unit[0]{$^{\circ}}$ with respect to the laser polarisation. Details
about the numerical simulations are discussed below. In
Fig.~\ref{fig:suppl_tdse_yield}\,(b) the photoelectron yield
integrated over the kinetic energy axis in
Fig.~\ref{fig:suppl_tdse_yield}\,(a) is shown. The maxima
and minima of the yield coincide with the extrema of the vector
potential.
The yield modulation is almost $\unit[20]{\%}$. In
Fig.~\ref{fig:suppl_tdse_yield}\,(c) the yield over the entire
detector (integration over all emission angles) is plotted. Here a
yield modulation is also observed, albeit of much smaller magnitude
($\approx\unit[2]{\%}$) and with a $T/2$ periodicity and always
increasing with respect to the unstreaked region of the
trace. Fig.~\ref{fig:suppl_tdse_yield}\,(d) shows the photoelectron
momentum distribution at large delay between NIR and XUV pulses, which
represents XUV only photoionisation. In
Fig.~\ref{fig:suppl_tdse_yield}\,(e) the momentum distribution is
shown, corresponding to a minimum in the amplitude of the vector
potential of the NIR field. Here the momentum spheres are shifted
along the polarisation axis. Also a change in the yield is observable,
with an increased number of photoelectrons emitted in the direction
towards which the momentum sphere is shifted. Both the
small (Fig.~\ref{fig:suppl_tdse_yield}\,(c)), and the large
modulations in the photoelectron yield
(Fig.~\ref{fig:suppl_tdse_yield}\,(a), (e)) can be explained by higher
order processes in the interaction of the few-cycle strong NIR pulse
in combination with the XUV field with the neon target atom.  Due to
the strong NIR field and very short pulse duration one can expect a
high degree of ground state polarisation, which in turn strongly
alters the angle-dependent photoionisation probability.

To study the influence of the aforementioned effects on the
retrievals, the VTGPA and SGO methods were applied to simulated
electron streaking traces. The simulated streaking traces were
obtained by solving the time dependent Schr\"{o}dinger equation (TDSE)
using the SCID code, described in~\cite{patchkovskii_simple_2016}. The
single active electron neon potential used to describe the different
ionisation channels is described in~\cite{Serguei_Ne}. The spatial
simulation box consisted of 4000 points, with a step-size of
$\unit[0.05]{a.u.}$. Angular momenta up to $l\leq40$ were included in
the angular part of the wave-function decomposition. In order to avoid
reflections from the edges of the simulation box, a complex absorber
potential~\cite{manolopoulos_derivation_2002} was placed at a distance
of $r= \unit[167.3]{a.u.}$ from the origin. The photoelectron spectra
for each of the time delays were calculated using the iSURFV
method~\cite{morales_isurf_2016}, where the matching distance was set
at a distance of $r=\unit[157.3]{a.u.}$ from the origin.
    
Electron streaking traces at different observation angles were
constructed from the angle-resolved photoelectron momentum
distributions calculated for varying XUV-NIR delays. In order to
recreate conditions similar to the experiment, one pulse retrieved
from one experimental data set was utilised as the XUV pulse in the
TDSE calculations. The NIR pulse used in the simulations was an
\unit[800]{nm}, \unit[5]{fs} FWHM transform-limited pulse with a
Gaussian envelope and an intensity of \unit[5]{TW/cm$^2$}.
Given that the XUV spectrum used for the TDSE simulations is known,
the absolute value of the energy- and angle-dependent dipole matrix
element that enters equation~(1) in the main manuscript can be
extracted from XUV-only TDSE results. The dipole term constructed in
this way was used for both the retrievals from simulated streaking
traces, and the pulse retrievals from experimental data. The phase of
the dipole matrix element was ignored, which is a reasonable
approximation given that neon has no resonances in the continuum in
the spectral region under consideration. A more detailed discussion
follows below.

\begin{figure}
  \centering
  \includegraphics[width=\textwidth]{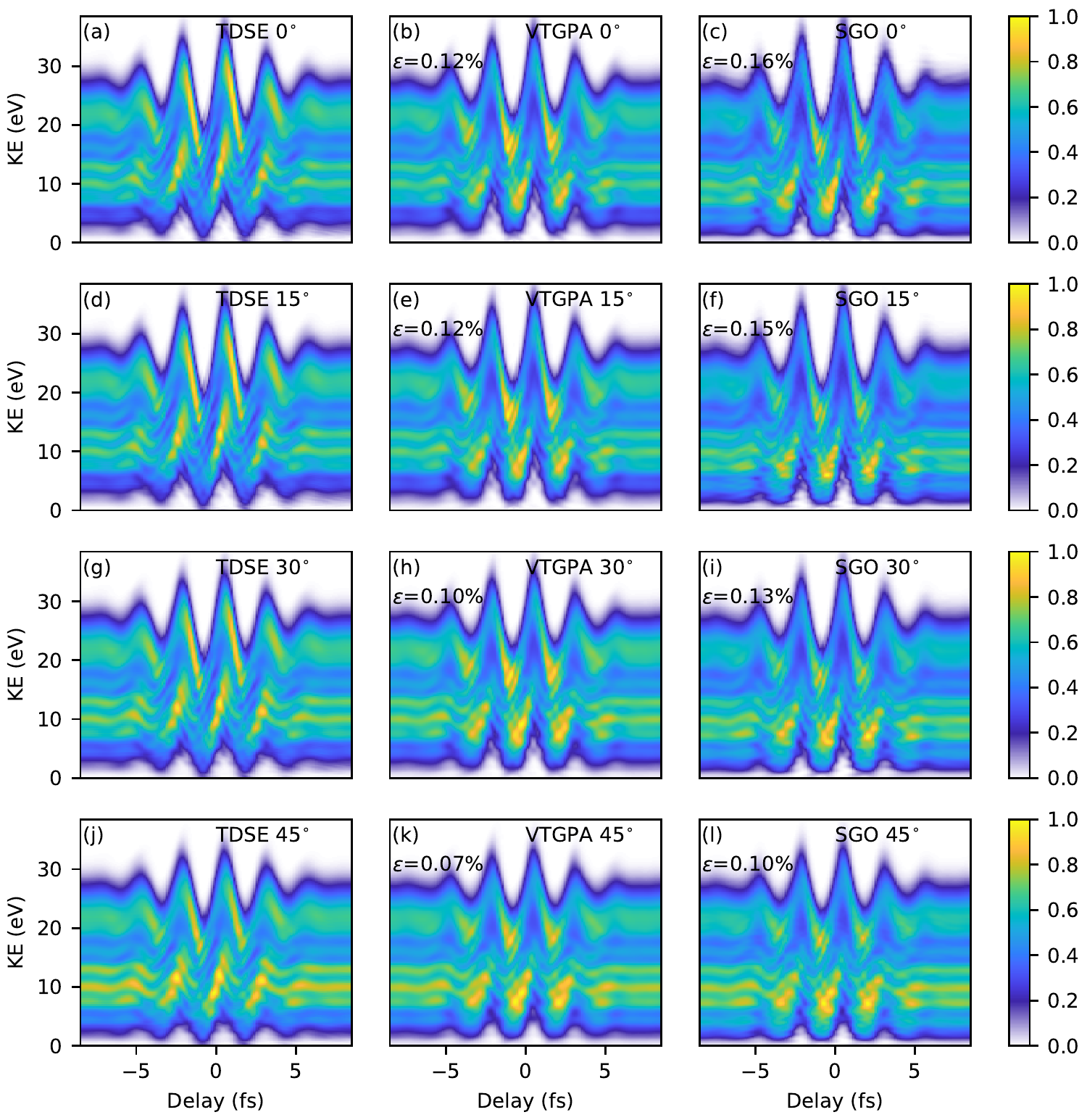}
  \caption{\textbf{$\mid$ Simulated and retrieved streaking
      traces}. Left column: simulated traces at different observation
    angles; (a) 0\,$^{\circ}$, (e) 15\,$^{\circ}$, (h) 30\,$^{\circ}$
    and (k) 45\,$^{\circ}$. Center: corresponding reconstructed traces
    by the VTGPA ((b), (f), (i) and (l) respectively. Right column:
    corresponding reconstructed traces by the SGO ((c), (g), (j)
    and (m) respectively).}
\label{fig:suppl_traces_tdse_vtgpa}
\end{figure}

\begin{figure}
  \centering
  \includegraphics[width=\textwidth]{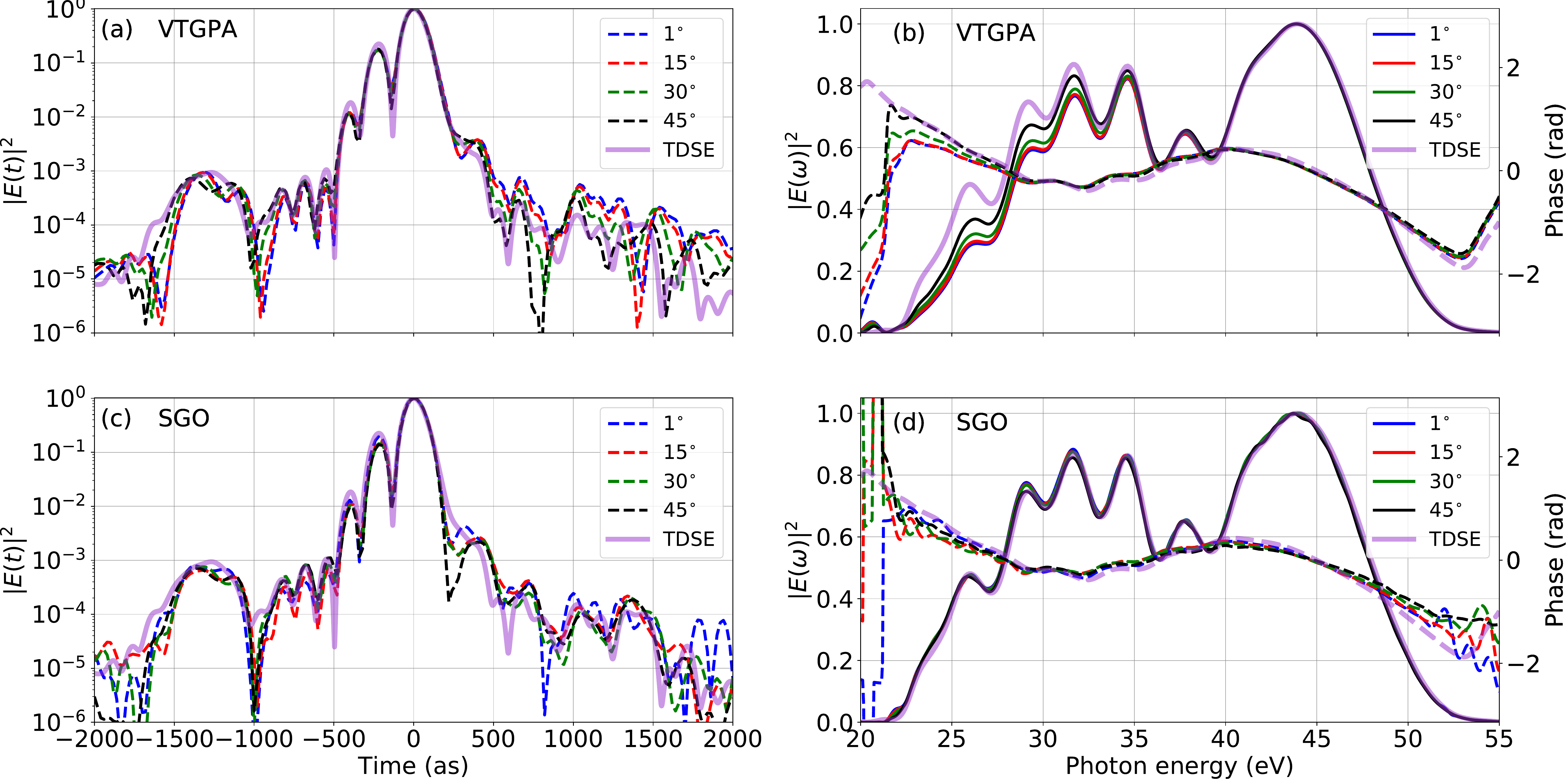}
  \caption{\textbf{$\mid$ Retrieved pulses and spectra from simulated
      traces.} (a) Retrieved pulse temporal shape on logarithmic scale
    from each retrieval at different observation angles with the VTGPA
    method in comparison with the original input pulse. (b)
    Corresponding spectra and spectral phases. (c) and (d) show the
    retrieved pulses in time and frequency domains respectively for
    the retrievals performed with the SGO method.}
 \label{fig:suppl_pulses_tdse}
\end{figure}

Fig.~\ref{fig:suppl_traces_tdse_vtgpa} shows the
simulated (left column), retrieved with the VTGPA (center column) and
with the SGO (right column) electron streaking traces, at four
different observation angles: (a), (b), (c) 0\,$^{\circ}$, (e), (f),
(g) 15\,$^{\circ}$ , (h), (i), (j) 30\,$^{\circ}$, and (k), (l), (m)
45\,$^{\circ}$. All retrieved traces include the figure of merit
$\epsilon_{\text{trace}}$.
%
The retrieved traces reproduce most features of the simulated traces
qualitatively, apart from a set of barely visible (see
Fig~\ref{fig:suppl_traces_tdse_vtgpa} (a),(d),(g),(j)) hyperbolic
interference fringes at low energy and positive delays, which are the
result of the coherent addition of direct ionisation by the XUV, and
excitation of a bound state by the XUV followed by ionisation by the
NIR field~\cite{mauritsson_2010}.
From
Fig.~\ref{fig:suppl_traces_tdse_vtgpa} one can anticipate
that the retrieved NIR intensity has been overestimated by the SGO
algorithm. The retrieved intensities for the VTGPA retrievals were
\unit[5.0]{TW/cm$^2$}, \unit[5.3]{TW/cm$^2$}, \unit[5.3]{TW/cm$^2$},
and \unit[4.8]{TW/cm$^2$} for observation angles of 0\,$^{\circ}$,
15\,$^{\circ}$, 30\,$^{\circ}$, and 45\,$^{\circ}$ respectively and in
the case of the SGO the retrieved intensities were
\unit[6.0]{TW/cm$^2$}, \unit[6.0]{TW/cm$^2$}, \unit[6.0]{TW/cm$^2$},
and \unit[5.8]{TW/cm$^2$}.
Results of the retrievals with both methods at the previously
mentioned four different observation angles, and comparison with the
input XUV pulse are presented in
Fig.~\ref{fig:suppl_pulses_tdse}.
The retrievals with the VTGPA Fig.~\ref{fig:suppl_pulses_tdse}\,(a)
and the SGO Fig.~\ref{fig:suppl_pulses_tdse}\,(c) methods reproduce
the shape of the original XUV pulse very well, including the satellite
pulses with intensities around \unit[0.1]{\%} of the maximum
intensity. The intensity of the pre-pulse at approximately
\unit[-1.35]{fs} is underestimated by an approximate factor of 0.7 in
the SGO retrievals, while the VTGPA retrievals on average
underestimate the intensity of the pre-pulse by a factor of 0.9. In
the case of the post-pulse at approximately \unit[1.35]{fs} with an
intensity of \unit[0.01]{\%} of the maximum, the SGO retrievals
overestimate the intensity of the peak on average by a factor of 1.1,
while the VTGPA retrievals do not reproduce this peak in the
reconstructed temporal pulses. In the experimental data, these
satellite pulses would correspond to harmonic generation by the two
half-cycles adjacent to the main half-cycle of the NIR driving laser.


The main features of the main pulse are successfully retrieved by
both, the VTGPA and the SGO methods. The average retrieved pulse
duration (full-width at half maximum of the intensity envelope) by the
VTGPA is \unit[140$\pm$2]{as}, compared to the \unit[134]{as} pulse
duration of the pulse used in the TDSE. Meanwhile, the average pulse
duration retrieved by the SGO is \unit[136$\pm$2]{as}. The shoulder on
the leading edge of the main pulse due to a third-order term in the
spectral phase is well reproduced for all retrievals, although both
algorithms seem to slightly underestimate the magnitude of the third
order dispersion. This is also observable in the spectral domain
plot. Based on the simulations we conclude that the retrieved pulses
reproduce the original pulse with high fidelity down to intensities of
\unit[0.1]{\%} of the maximum.

In order to quantify the differences between the retrieved and
original pulses, the root-mean-square field (RMSF) error defined by
Dorrer and Walmsley~\cite{Dorrer_rmsf} was determined for all
retrievals. The RMSF error is defined according to:
\begin{equation}
\epsilon_{\text{RMSF}} =
\int_{-\infty}^{+\infty}|E_{\textrm{TDSE}}(t)-E_{\textrm{ret}}(t)|^2
\,dt\, ,
\label{eq:rmsf}
\end{equation}
where $E_{\textrm{TDSE}}(t)$ is the input XUV pulse used in the TDSE
calculations and $E_{\textrm{ret}}(t)$ is the pulse retrieved either
by the VTGPA or the SGO method. In order to compute a dimensionless
quantity both fields are normalized according to:
\begin{equation}
E_{\textrm{normalized}} = \frac{E(t)}{\int_{-\infty}^{+\infty}|E(t)|^2 \,dt}
\label{eq:norm}
\end{equation}
The RMSF error defined in this way varies between 0 and 2, where a
number $\lesssim 0.1$ is considered to be an adequate pulse
retrieval. The calculated RMSF errors were $1.43 \times 10^{-1}$,
$1.39 \times 10^{-1}$, $1.16 \times 10^{-1}$, and
$0.87 \times 10^{-1}$ for the VTGPA results at observation angles of
0\,$^{\circ}$, 15\,$^{\circ}$, 30\, $^{\circ}$, and 45\,$^{\circ}$
respectively. Meanwhile, the calculated RMSF errors for the SGO
results were $0.83\times{}10^{-1}$, $1.30 \times 10^{-1}$,
$1.31 \times 10^{-1}$, and $1.27 \times 10^{-1}$ at observation angles
of 0\,$^{\circ}$, 15\,$^{\circ}$, 30\, $^{\circ}$, and 45\,$^{\circ}$
respectively. The results show that the solutions of the SGO at small
observation angles approximate the original pulse slightly better than
the solutions of the VTGPA. At the two lowest observation angles
however, the VTGPA yields a smaller RMSF. Comparing the trace errors
$\epsilon_{\text{trace}}$ to the pulse field errors
$\epsilon_{\text{RMSF}}$ it is difficult to establish a direct
correlation between these two errors. This highlights the challenge to
find a unique solution in the reconstruction problem discussed at the
end of section~\ref{attopulseretrievalalgorithm}.

Finally, the influence of the phase term in the complex dipole matrix
element was tested. To this end, retrievals for streaking traces at an
observation angle of 45\,$^{\circ}$ were performed, but modifying the
dipole matrix element in eq.~(1) of the main manuscript by adding an
energy-dependent phase according to the results published in
Mauritsson~et~al.~\cite{mauritsson_accessing_2005}. The retrieved
spectra, spectral phases, and temporal shapes (not shown) were hardly
affected. In the VTGPA retrieval the RMSF without phase is
$\unit[0.87 \times 10^{-1}]{}$; with dipole phase the RMSF is
$\unit[0.89 \times 10^{-1}]{}$. For the SGO retrievals the RMSF is
$\unit[1.27 \times 10^{-1}]{}$ without dipole phase, and
$\unit[1.09 \times 10^{-1}]{}$ with dipole phase.
These changes in the RMSF of the reconstructed pulses are on the order
of magnitude of the variations that are observed for retrievals at
different observation angles. Therefore, we conclude that the
inclusion or not of an energy-dependent phase into the dipole matrix
element does not play a significant role for our retrievals. In view
of the previous observations we conclude that the VTGPA and SGO
methods are suitable for retrievals of experimental data at different
observation angles.  From the results of the simulated data and
corresponding retrievals, we conclude that the influence of the
Coulomb field, ionisation through intermediate states, and modulations
of the electron yield do not hinder the pulse retrieval.

\section{Attosecond pulse retrievals from experimental traces}

\begin{figure}
  \centering
  \includegraphics[width=\textwidth]{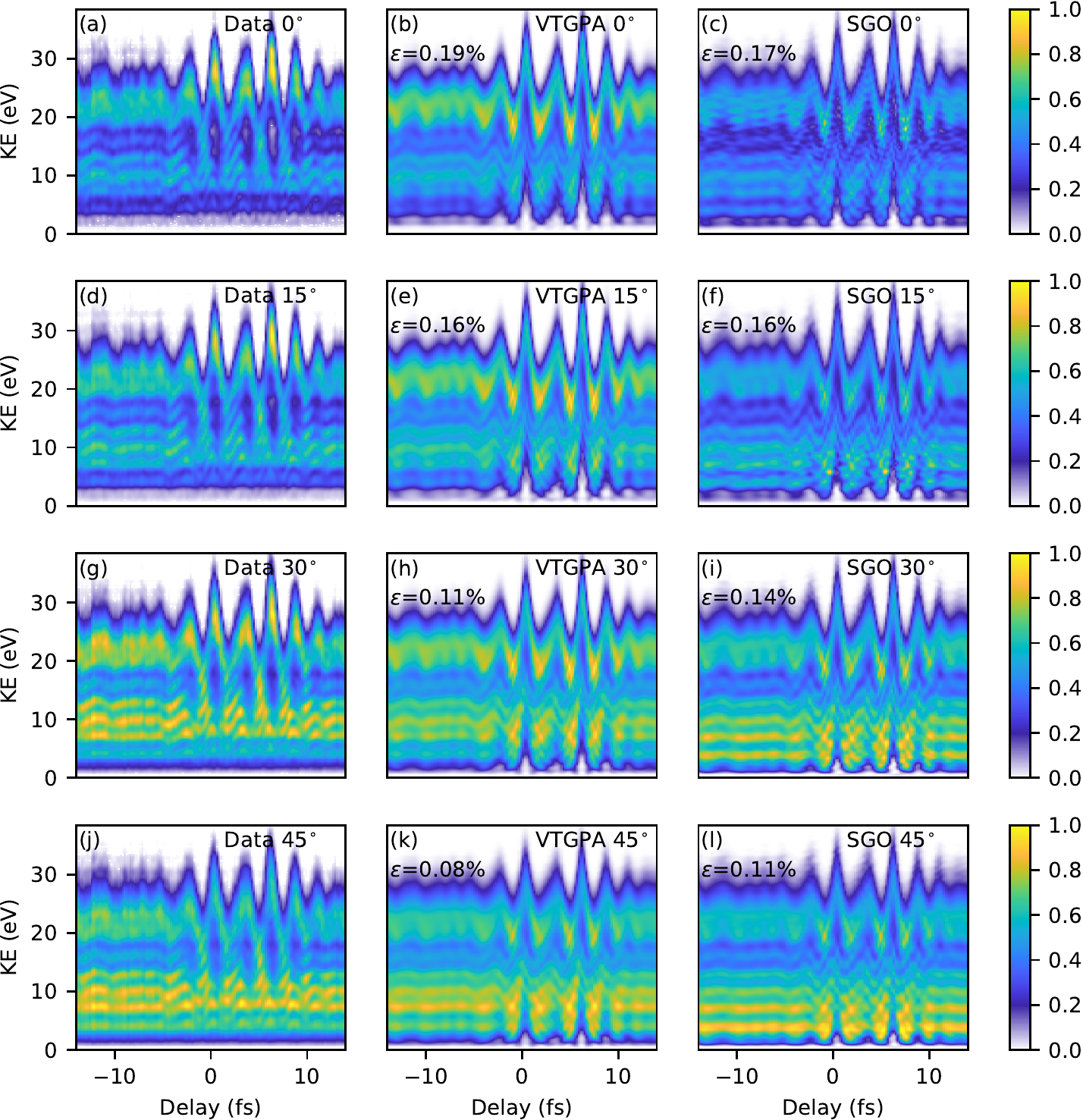}
  \caption{\textbf{$\mid$ Experimental and retrieved streaking
      traces.}  left column: measured attosecond electron streaking
    traces at the observation angles (a) $\theta$ =
    0\,$^{\circ}$, (d) $\theta$ = 15\,$^{\circ}$, (g) $\theta$ =
    30\,$^{\circ}$, (j) $\theta$ = 45\,$^{\circ}$. Center column:
    corresponding retrieved traces with the VTGPA method: (b) $\theta$
    = 0\,$^{\circ}$, (e) $\theta$ = 15\,$^{\circ}$, (h) $\theta$ =
    30\,$^{\circ}$, (k) $\theta$ = 45\, $^{\circ}$. Right column: the
    corresponding retrieved traces with the SGO method.}
\label{fig:suppl_traces_vtgpa}
\end{figure}

The retrievals were carried out for experimental streaking traces
obtained from the Abel-inverted experimental images at eight different
observation angles 0\,$^{\circ}$, 15\,$^{\circ}$, 30\,$^{\circ}$,
45\,$^{\circ}$, 135\,$^{\circ}$, 150\,$^{\circ}$, 165\,$^{\circ}$, and
180\,$^{\circ}$ with respect to the polarisation direction.
For each of the eight resulting individual streaking traces, four
different post-processing methods were applied: background
subtraction, background subtraction and correction of the
delay-dependent total electron yield, background subtraction and
fast-Fourier transform filtering (FFT filtering), and background
subtraction plus yield correction plus FFT filtering. For the yield
correction, the total yield was integrated in each half (left/right of
the polarisation vector) of the experimental VMI images at each delay,
to account for a slight drift of the total electron yield over
time. The VTGPA and the SGO algorithms were applied to each of the 32
resultant experimental traces. Fig.~\ref{fig:suppl_traces_vtgpa} shows
a set of experimental traces with background subtraction plus FFT
filtering at observation angles of (a) 0\,$^{\circ}$, (d)
15\,$^{\circ}$, (g) 30\,$^{\circ}$, and (j) 45\,$^{\circ}$, and the
corresponding retrieved traces with the VTGPA ((b) 0\,$^{\circ}$, (e)
15\,$^{\circ}$, (h) 30\,$^{\circ}$, and (k) 45\,$^{\circ}$), and the
SGO ((c) 0 \,$^{\circ}$, (f) 15\,$^{\circ}$, (i) 30\,$^{\circ}$, and
(l) 45\,$^{\circ}$).
Each of the figures featuring reconstructed streaking traces in
Fig.~\ref{fig:suppl_traces_vtgpa} includes a relative error calculated
according to eq.~\eqref{eq:err}.

\begin{figure}
  \centering
  \includegraphics[width=\textwidth]{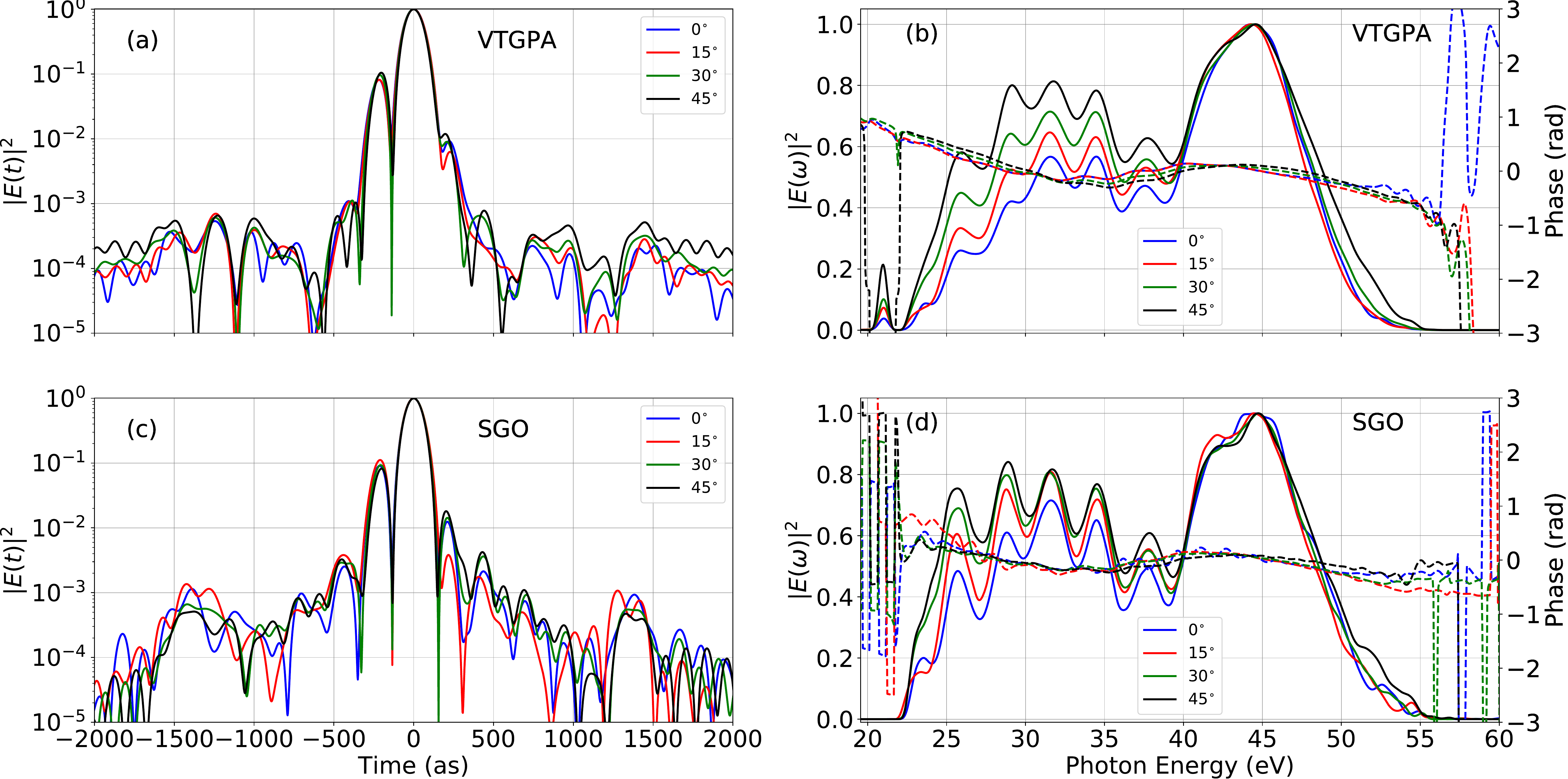}
  \caption{\textbf{$\mid$ Retrieved pulses and spectra from
      experimental traces.} Left column: temporal intensity profiles
    of retrieved pulses on logarithmic scale from each retrieval at
    different observation angles for VTGPA (a) and SGO (c). Right
    column: corresponding retrieved spectra (solid lines) and spectral
    phases (dashed lines) for VTGPA (b) and SGO (d). In all figures,
    the results from data at different observation angles are
    represented with different colors; blue: $\theta$ = 0\,$^{\circ}$,
    red: $\theta$ = 15\, $^{\circ}$, green: $\theta$ = 30\,$^{\circ}$,
    black: $\theta$ = 45\,$^{\circ}$.}
 \label{fig:suppl_retrieved_pulses}
\end{figure}

Fig.~\ref{fig:suppl_retrieved_pulses} shows the reconstructed fields
in the time and frequency domain corresponding to the traces in
Fig.~\ref{fig:suppl_traces_vtgpa}. Both, the VTGPA and the SGO
algorithms retrieve a dominant main pulse with negative TOD. Both
algorithms also retrieve satellite pulses corresponding to XUV
emission from adjacent half cycles of the driving NIR pulse. The SGO
algorithm retrieves slightly stronger satellites. The satellites have
intensities at or below $\unit[0.1]{\%}$ of the main XUV pulse. The
retrieved spectra have a smooth component in the high energy portion,
and exhibit fringes for energies below \unit[40]{eV}. These fringes
stem from interference between the satellites and the main pulse with
the fringe spacing of $\approx\unit[3.3]{eV}$ corresponding to half
the period of the driving NIR field with a centre wavelength of
$\lambda_0 = \unit[760]{nm}$.

\clearpage
\section*{Supplementary Information References}